\documentclass{epl}
\usepackage{graphicx}

\title{Ratchet Cellular Automata for Colloids in Dynamic Traps}  
\shorttitle{Ratchet Cellular Automata for Colloids...}
\author{C.J. Olson Reichhardt and C. Reichhardt} 
\institute{
Center for Nonlinear Studies and Theoretical 
Division, 
Los Alamos National Laboratory, Los Alamos, New Mexico 87545 USA}
\begin{document}
\maketitle

\begin{abstract}
We numerically investigate the transport of kinks in 
a ratchet cellular automata geometry for colloids interacting with
dynamical traps. We find that thermal effects can enhance the transport
efficiency in agreement with recent experiments. At high temperatures 
we observe the creation and annihilation of thermally induced kinks    
that degrade the signal transmission. We consider both the
deterministic and stochastic cases and show how the trap geometry can 
be adjusted to switch between these two cases. 
The operation of the dynamical trap geometry can be achieved
with the adjustment of fewer parameters than  
ratchet cellular automata constructed using static traps. 
\end{abstract}

Recently, a ratchet mechanism was proposed
for propagating logic states in a clocked manner 
through a system of  vortices in nanostructured type-II superconductors  
\cite{Hastings}.  Since the operation of the device depends on the
discrete positions of the vortices,
the system was termed a ratchet cellular automata (RCA).
It has been demonstrated that a complete logic architecture
can be constructed using the RCA, so that variations of RCAs
constructed in different systems might offer a promising 
alternative to the current microelectronic 
logic architectures based on silicon MOSFETs \cite{Hastings,Ratchet}. 
The use of discrete particles to store logic states or perform
logic operations has been studied previously in various forms 
including the quantum dot cellular automata \cite{Lent} 
and magnetic dot cellular automata \cite{Cowburn}.  
In contrast to these systems, which work in the adiabatic limit,
an RCA operates when the system is 
far from equilibrium.   

The original RCA geometry was proposed for 
vortices in a type-II superconductor with
nanostructured pinning sites.  The vortices act like repulsive 
particles and adopt one of two possible configurations in the static
pinning sites.
In order to propagate a logic signal through the device,
an alternating external driving force must be applied, such as
by inducing an oscillating Lorentz force on the vortices by means of
an alternating current. 
The basic concept of the RCA 
should be applicable to any system of particles which have a repulsive
interaction with each other,
such as ions in optical traps, classical electrons, 
vortices in Bose-Einstein condensates with 
optical arrays of traps \cite{Bigelow}, or colloids 
interacting with arrangements of optical traps \cite{Babic}. 
In many of these systems, such as for the colloids, 
it is already experimentally possible
to create dynamical traps; in this case, an external applied drive may not
be necessary. 

In the original RCA geometry, the basic structural unit consists of three 
elongated static traps, each containing a single vortex. 
The three traps have different widths and biases in order to break the spatial
symmetry of the system.  In each trap, the vortex sits either at the top
or the bottom of the trap and represents either state 0 or 1.
A three-stage alternating external drive is then applied which shifts
the vortices to the left and right inside the traps.  The trap designs are
chosen such that the result of these shifts is to alter the distance
between vortices in neighboring traps  
so that two out of every three vortices
are close to
one neighboring vortex and far from the other.  The resulting
asymmetry permits propagation of the logic signal, which would otherwise
not occur in this overdamped system.
With this geometry, a kink corresponding to a change in logic state can
be propagated along a chain of traps.
The bare RCA functions at finite temperatures in a stochastic mode since
a small barrier remains at the center of each trap which must be overcome
thermally.  If an additional potential is superimposed on the wells to
counteract this barrier, the RCA can operate in a completely deterministic
mode and can also run at $T=0$.

Recently, an experimental version of RCA has been realized for colloids 
confined to two dimensions and
interacting with
optical traps \cite{Bechinger,Babic}. The colloids are 
micron-sized particles with 
repulsive Yukawa or screened Coulomb interactions. 
In this case, a series of optical traps are prepared and the wells are
labeled as sites A, B, and C in a pattern that repeats across all of the wells.
Each trap is composed of a double well potential created by two optical
tweezers, so the colloidal particles sit either in the up or down position
inside each trap.  Unlike the original RCA, in the colloidal realization each 
trap is identical in shape.  
The ratchet effect is induced by dynamically relocating the positions
of the wells 
periodically.  The wells labeled B are moved
to the right and left, and separately the wells labeled C are also moved
to the right and left, so that the net effect is the same alteration of
spacing between neighboring colloids that was achieved by means of three
well shapes and a three-stage alternating drive in the vortex system
of Ref.~\cite{Hastings}.

For example, a colloid in well B1 is initially close to the colloid in well
A1 and far from the colloid in well C1, 
shown in Fig.~1.
When the position of the colloid in
well A1 is switched from the up to the down position, then the position of 
the colloid in well B1 flips from the down to the up position.  The signal
does not propagate any further until the wells in group C are moved to the
new position marked by dashed lines in Fig.~1 and labeled (B/C).  This
brings colloids B1 and C1 close together while placing colloids C1 and A2
far apart, so that colloid C1 flips to the new state.
In the next stage, the wells in groups B and C are simultaneously shifted
to the right so that the wells in group C occupy their original positions
while the wells in group B occupy the position labeled (B/C).  This
brings colloids C1 and A2 close together while moving B2 far from A2, and 
permits colloid A2 to switch.
The wells in group B are moved back to their original positions and
the process is repeated with
the change in colloid position propagating as a dislocation which follows
the far spacing of the wells.
 
The colloidal version of the RCA which has been experimentally realized 
will be a useful system for studying alternative geometries and 
further properties of the RCA.   
The colloid version of RCA can also provide a valuable system with which
to understand 
transport in noisy environments, which has connections to stochastic resonance.
Several important issues have not been studied directly
in the experiments, such as explicitly changing
the temperature or the clock frequency, 
as well as understanding the role of thermally 
induced kink and anti-kink creation in the signal propagation.
It would be valuable to probe
the effect of trap geometry on the transition
between the thermally dominated and deterministic or clocked regimes.
Here, we explore all of these possibilities through simulations.

\begin{figure}
\includegraphics[width=3.5in]{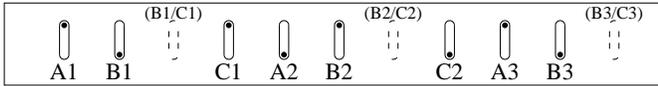}
\caption{
Image of the system showing colloid positions (black dots)
inside the three sets of wells marked $A$, $B$ and $C$ at the
stage of the ratchet cycle where the far spacing is between
wells $B$ and $C$. 
Wells $A$ are stationary, while the wells in sets
$B$ and $C$ move back and forth to the dashed locations.
}
\end{figure}

{\it Simulation:}
We consider a quasi one-dimensional geometry of $N_c=144$ traps 
with open boundary conditions,
in analogy with the geometry considered 
in the experiments of Ref.~\cite{Babic}.
Each trap contains a single colloid which is
modeled as an overdamped particle 
confined to two dimensions and interacting
with the other colloids via a Yukawa potential 
$U(r_{ij}) = q^2\exp(-\kappa r_{ij})/r_{ij}$.  
Here $\kappa=1$ is the inverse screening length and $q$ is 
the colloid charge measured in dimensionless units.
The equation of motion for a colloid $i$ 
is
\begin{equation}
\eta{\bf v_i} = {\bf F}_{i} =  
{\bf F}_{i}^{cc} + {\bf F}_{trap}^i +{\bf F}_i^{T} .
\end{equation}
Here  
$\eta$ is the damping constant 
from the surrounding fluid which is set to unity. 
The colloid-colloid force 
${\bf F}_{i}^{cc} = -\sum^{N_c}_{j \neq i}\nabla_i  U(r_{ij})$. 
Since the colloid interaction force falls off exponentially for
large $r$ we place a cutoff on the interaction 
at $r=4$, further than the screening length, for computational
efficiency. 
Taking a longer cutoff produces the same results.
The temperature is applied as random Langevin kicks
${\bf F}^{T}$ 
with the 
statistical properties $\langle f^{T}(t)\rangle = 0$ 
and $\langle f^{T}(t)f^{T}(t^{\prime})\rangle 
= 2\eta k_{B}T\delta(t -t^{\prime})$. 
The trap force 
${\bf F}_{trap}$ is produced by lozenge shaped
pins, each of which is composed of 
two half-parabolic traps separated by an elongated
region that confines only in the $x$ direction.
The aspect ratio of the pins is 3 to 1, with the
long direction running along the $y$ axis perpendicular to the line
of pins, as in Fig.~1.  The pinning strength $f_p=11.0$.
The lozenge shapes of the traps were chosen to model the
experiments closely.
The positions of the wells in group A is fixed in time, and the
distance between the wells in group A, which corresponds to the
lattice constant of the ratchet device, is 5.0.  
The wells B and C are moved back and forth periodically in three
stages.  In the first stage, both wells B and C are shifted to the right
by a distance of 1.0.  In the second stage, 
the wells in group
B are moved back by a distance of 1.0 to the leftmost position.  In the
third stage, 
the wells in group C are moved back by a distance of 1.0
to the leftmost position.  The cycle then repeats beginning with both
wells B and C in their leftmost positions.
The total length of time spent by the wells in one of
the three stages is reported as the clock period $\tau$.
A kink in the form of a change in logic state is produced by
moving the leftmost colloid from the up to the down position
at time $t=0$.
The system operates stochastically due to the presence of a finite barrier
at the center of each well, generated by interparticle interactions,
and at $T = 0$
there is no transmission of kinks. At finite $T$ the kink propagates 
and the
system can exhibit either a clocked or deterministic behavior.    

\begin{figure}
\includegraphics[width=3.5in]{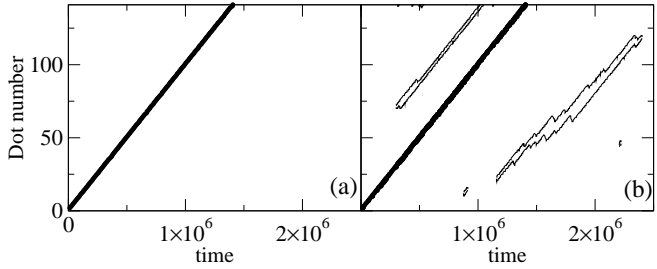}
\caption{
(a) The kink position vs time
for a system with $\tau=10000$ at $T = 0.4$. (b) 
The same system at $T = 1.15$.  Dark dots indicate propagation of the
originally introduced kink, while light dots show the formation and
propagation of thermally activated kink-anti-kink pairs.
}
\end{figure}

In Fig.~2(a) we plot the location of
a kink that was inserted at the edge of the sample
at $t = 0$ as a function of time for $T = 0.4$    
in a system with $q^2=0.5$ and $\tau=10000$.
At this temperature, the kink moves in a 
clocked manner through the entire system of 144 dots, and
there are no thermally created kinks or anti-kinks.
The kink propagates at a constant speed, 
as indicated by the linear
slope.  
We observe a similar clocked
motion at lower temperatures until $T < 0.1$. 
Below this temperature the kink becomes pinned near its entry point
and does not propagate across the system.
For $T = 1.15$ for the same system, shown in Fig.~2(b), 
thermally
induced kinks can appear. The initial kink is marked as a thick black line,
and moves through the system 
at the same speed as the kink in Fig.~2(a). 
At later times, thermally activated kink-anti-kink pairs are
created, and the ratcheting mechanism
propagates both species of kinks in the same direction. 
Kinks and anti-kinks collide and annihilate
when the leading kink takes a thermally induced step backwards 
and the anti-kink is able to catch it. 
In other cases the kinks and anti-kinks travel the length of  
the system without annihilating.  
  
\begin{figure}
\includegraphics[width=3.5in]{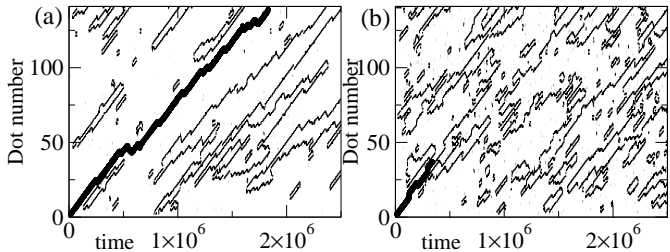}
\caption{
The same system as in Fig.~2 for (a) $T = 1.3$;
(b) $T = 1.4$.  
Dark dots: the originally introduced kink; light
dots: thermally activated kink-anti-kink pairs.
}
\end{figure}

In Fig.~3(a), we show the same system 
at $T = 1.3$. In this case the
thermal fluctuations are sufficiently strong that the 
induced kink does not move linearly  but shows occasional steps backward
so the motion is no longer completely clocked. 
A significant number of thermally induced kink anti-kink pairs form
and also show occasional steps backwards. 
As the temperature increases the average number of thermally 
created pairs 
increases and the average lifetime of a given pair decreases.
Figure 3(b) illustrates the system at $T = 1.4$, 
where thermally created pairs
proliferate rapidly and the initially introduced kink is 
both hard to distinguish and short-lived. 
For $T > 1.4$ the thermal fluctuations are so strong that 
the colloids can hop out of the individual wells.  

\begin{figure}
\includegraphics[width=3.5in]{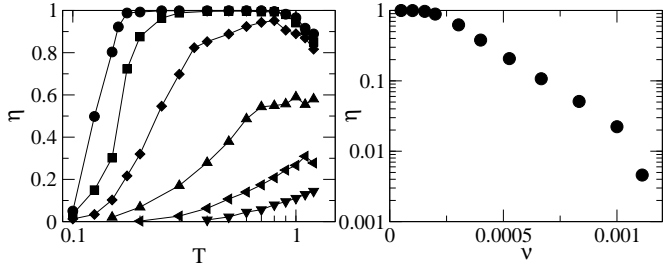}
\caption{ 
(a) The efficiency $\eta$ vs 
$T$ for $q^2=0.5$ at clocking speeds 
$\tau=$ 20000, 10000, 5000, 2500, 1500, and 1000 from top to bottom.
(b) $\eta$ vs clock frequency $\nu$
in inverse MD steps for fixed $T = 0.5$. 
}
\end{figure}

The kink motion can also be characterized by measuring 
the transmission efficiency $\eta$, which is defined 
in terms of the time $\tau_{det}$ it would 
take for a kink in the completely deterministic system to travel the
length of the system. 
The actual travel time for a kink is given by $\tau_{kink}$, so
the efficiency $\eta = \tau_{kink}/\tau_{det}$.
If the kink travels at the clocked pace, $\eta=1$. If the kink 
takes steps backward, gets stalled, or annihilates with a
thermally activated anti-kink, $\eta<1$.
In Fig.~4(a) we plot $\eta$ vs $T$ for 
fixed $q^2 = 0.5$, $f_{p} = 11.0$,     
and different frequencies or clocking speeds. 
Each point has been averaged over five realizations of thermal
disorder.
In this system the kinks are 
motionless for $ T < 0.1$ since the particle-particle interactions induce a
barrier
at the center of the pinning sites that must be overcome by
a small amount of thermal activation.
For $T\ge 0.1$, there are enough thermal fluctuations to overcome
the barrier and the kinks begin to propagate.  
For 
$\tau > 10000$, 
kinks can propagate deterministically, 
and $\eta=1$ over a wide
range of temperatures.
The efficiency begins to drop when $T>0.8$ 
since there are excessive
thermal fluctuations that cause the kinks to take occasional steps
backward rather than forward.
In the experiments of Ref.~\cite{Babic}, 
there was also a certain range of parameters over which the
system operated deterministically and $\eta=1$.
For shorter clock periods, $\tau_{det}$ decreases and
the kink should in principle move through the chain at a faster rate.
Instead, as the clock frequency increases,
the colloids are no longer able to respond    
and the efficiency decreases. Fig.~4(a) shows that 
as the clock period decreases,
the temperature at which the system reaches a deterministic 
mode increases.  For 
$\tau<10000$
the system is never able to enter the fully deterministic region.
There is an upper bound on the temperature that can be applied to this
system, since for $T>1.4$ the colloids begin to jump completely out of the
wells by thermal activation and the ratchet device is destroyed.
In Fig.~4(b) we show the efficiency vs clock 
frequency $\nu$ for fixed $T = 0.5$. 
There is a plateau of $\eta=1$
at low clock frequencies followed by an exponential decrease 
of $\eta$ with increasing $\nu$
which is cut off at the highest frequencies.

\begin{figure}
\includegraphics[width=3.5in]{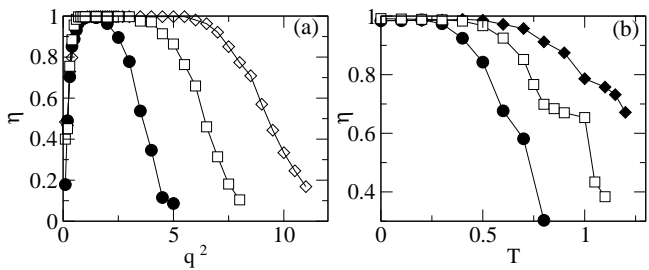}
\caption{
(a) Efficiency vs $q^2$ for constant $T$ and fixed $\tau=5000$.
Diamonds: $T=1.0$.  Open squares: $T=0.75$.  Filled circles: $T=0.5$.
(b) Efficiency vs $T$ for elliptical traps at $\tau=5000$.  
Filled circles: $q^2=18$.  Open squares: $q^2=18.5$.  Filled diamonds:
$q^2=19.$
}
\end{figure}

In the experimental work of Ref.~\cite{Babic}, 
it was argued that increasing
the strength of the central trap in each pin has an effect that is 
similar to decreasing the temperature. 
This implies that if the central trap strength is
fixed, then there should be an optimal temperature range for deterministic
transport of logic signals.
The results in Fig.~4 support this conclusion. 
The effect of temperature is more pronounced at the lower clock periods
$\tau$ and the temperature window in which deterministic behavior occurs
shrinks with 
decreasing $\tau$.
In the case of $\tau=5000$ in Fig.~4(a),
there is a peak value of $\eta$ near 
$T = 0.9$.  

In our model, varying the trap strength $f_p$ does not affect the 
ratchet efficiency.  This is because it is the interaction
forces between the colloids that control both the strength of the 
induced barrier
at the center of a pin as well as the magnitude of the force that leads
to propagation of the logic signal from pin to pin.  We have tested both
larger and smaller values of $f_p$ and find that the 
pinning can be made arbitrarily
strong without affecting our results.  In the experimental system, there is
a practical limit on the amount of laser power that can be provided to the 
sample in order to form the optical traps.  
There is also a physical limit to the amount of
energy that can be absorbed by the colloids and the bath medium over a given
time period before damage occurs.  For smaller values of $f_p$, which can 
be accessed readily in experiment
by reducing the laser power, we find that
the system is limited by the requirement
of keeping the colloids inside the pinning sites at all times.  
The colloids may depin if the colloid-colloid interaction force 
overcomes the pinning force, if thermal activation out of the pins
becomes possible, or if a combination of these two
effects occur.  Once the colloids depin, the ratchet is destroyed.

We have considered the effect of varying the strength of the colloid-colloid
interaction, $q^2$, as shown in Fig.~5(a) for different temperatures
at $\tau=5000$.
If $q^2$ is reduced toward zero, there is insufficient coupling between
the colloids for the ratchet mechanism to function, and $\eta$ drops
to zero.  As $q^2$ increases, the system enters the deterministic
regime with $\eta=1$.  At $q^2=0.5$, 
the same system with clock period $\tau=5000$ was shown in Fig.~4(a) never to
reach $\eta=1$ even as $T$ is increased.  
Fig.~5(a) indicates that for this clock speed, the
system can enter the deterministic limit if $q^2$ is increased to a value
of at least 1.  If $q^2$ is increased too much, however, the efficiency
drops again when the thermal fluctuations that are required
for operation of the ratchet are washed out by the very strong colloid-colloid
interaction forces.  We show the decrease in $\eta$ at higher values of
$q^2$ in Fig.~5(a).  At the lower temperature $T=0.5$,
$\eta$ drops below 1 once $q^2>1.6$, while for the higher temperature
$T=1.0$, thermal fluctuations are not washed out until $q^2>5.5$.

In the experiments of Ref.~\cite{Babic}, the traps used to construct the
ratchet had a multi-well shape which we have represented in our model by
a lozenge-shaped pin.  The same ratchet mechanism can also operate for
other types of wells.  To test this, we have considered a much simpler model
for the traps consisting of elliptical pins.  These are simply
parabolic traps with unequal aspect ratios in the $x$ and $y$ directions.
Unlike the lozenge-shaped pins, which have a central region that
is flat in the $y$ direction and 
confines only in the $x$ direction, the elliptical pins have a minimum
in both the $x$ and $y$ directions
at the center of the pin.  If the ratio of the pinning force to the
colloid-colloid interaction force is too small in the elliptical pin case,
the system loses its two logic states and all the colloids sit in the center
of the wells.  On the other hand, when the colloid-colloid interaction force
is strong enough that the alternating up-down configuration appears in the
elliptical pin
system, the ratchet mechanism can operate in a fully deterministic mode
even when $T=0$.  This is illustrated in Fig.~5(b) where we plot $\eta$ as
a function of $T$ for elliptical wells with $f_p=11.0$, 
$\tau=5000$, and $q^2=18$, 
18.5, and 19.
Here, we see that $\eta=1$ all the way down to and
including $T=0$, in contrast to the case in Fig.~4(a) where $\eta$ drops to
zero at $T=0$.  The central minimum in the elliptical pins compensates for
the potential barrier at the center of the pin induced by the colloid-colloid
interaction forces.  There is no benefit to adding temperature
to a system with elliptical pinning sites.  For low $T$, $\eta=1$, but as
$T$ increases, thermally activated kinks and antikinks appear and interfere
with the signal transmission, causing $\eta$ to drop.  Unlike the case
shown in Fig.~3 for the lozenge-shaped pins, where thermally activated kinks
tended to ratchet at nearly the clock speed through the system, in the
case of elliptical pins the thermally activated kinks tend to diffuse and are
much more likely to travel backwards than the kinks in the lozenge-shaped
pins.  
Fig.~5(b) also demonstrates that the transport efficiency of the 
elliptical traps at elevated temperatures is strongly
sensitive to the value of $q^2$.

We note that the RCA constructed using dynamical traps has several 
advantages over the originally proposed RCA which involved static traps 
in combination with an external drive. 
The static trap RCA requires three different
trap geometries to be constructed in the same system.
In order to 
achieve deterministic kink propagation
without temperature, an additional attractive
potential must be added to the center of each trap to compensate for 
the barrier in the middle of the trap caused by particle-particle 
interactions. 
Finally, a three stage external drive must be applied. In 
the dynamical trap RCA, 
many of these extra parameters are eliminated, making it easier to adjust
the system into a deterministic mode of operation. 
In the simplest case of elliptical traps, the system 
can function deterministically even without thermal fluctuations.
We also note that it may be possible to create
dynamical traps for vortices in type-II superconductors, the system in
which the RCA was first proposed.
It has been suggested recently that if artificial magnetic pinning sites
are created in a superconducting sample, 
then the strength and shapes of the pinning can be changed dynamically
by applying time dependent magnetic 
fields \cite{Carnerio}. 

In conclusion, we have numerically investigated ratchet cellular automata 
constructed for colloidal particles using dynamic traps. 
Our results are in good
agreement with the recent experiments of Ref.~\cite{Babic}. 
We considered the effects of changing several parameters that have
not been explored experimentally,
including temperature, colloid-colloid interaction strength, 
clock frequency, and the influence of
the trap geometry.
We find that 
temperature can  enhance the transport of kinks and can
permit the RCA to operate
in the deterministic 
limit even without the addition of an attractive potential to compensate
for the barrier created at the center of each pin by particle-particle
interactions. We also examined the proliferation of kink-antikink pairs   
which are created at higher temperatures. These  
pairs can be transported by the ratchet effect and 
can recombine and annihilate.  
At high clock frequencies, the efficiency of the ratchet transport is 
degraded since the colloids can no longer fully respond to the
ratchet mechanism. We also identify the existence of
optimal temperature regimes and particle interaction 
regimes for signal transport. 
In the case of elliptical traps, 
we show that the system can operate deterministically even in the
$T = 0$ limit.  

Acknowledgments---We thank D. Babic, C. Bechinger, and  M.B. Hastings 
for useful discussions. 
This work was supported by the US DoE under Contract No. W-7405-ENG-36.

\end{document}